\documentclass[%
reprint,
superscriptaddress,
showpacs,preprintnumbers,
nofootinbib,
 amsmath,amssymb,
 aps,
prb,
]{revtex4-1}

\usepackage{graphicx}
\usepackage{dcolumn}
\usepackage{bm}

\begin{document}

\title{Is there any hidden symmetry in the stripe structure of perovskite high temperature superconductors?}



\author{Vladimir A. Gavrichkov}
\affiliation{Kirensky Institute of Physics, Federal Research Center KSC Siberian Branch  of the Russian Academy of Sciences, 660036 Krasnoyarsk, Russia}
\author{Yury Shan'ko}
\affiliation{Institute of Computational Modeling of the Siberian Branch of the Russian Academy of Sciences, 660036 Krasnoyarsk, Russia}
\author{Natalia G. Zamkova}
\affiliation{Kirensky Institute of Physics, Federal Research Center KSC Siberian Branch of the Russian Academy of Sciences, 660036 Krasnoyarsk, Russia}

\author{Antonio Bianconi}
\affiliation{Rome International Center for Materials Science Superstripes (RICMASS), Via dei Sabelli 119A, 00185 Rome, Italy}
\affiliation{Institute of Crystallography, Consiglio Nazionale delle Ricerche, CNR, I-00015 Monterotondo, Italy}
\affiliation{National Research Nuclear University MEPhI (Moscow Engineering Physics Institute), 115409~Moscow,~Russia}



\date{\today}

\begin{abstract}
 Local and fast structural probes using synchrotron radiation have shown nanoscale striped puddles and nanoscale phase separation in doped perovskites.
 It is known that the striped phases in doped perovskites are due to competing interactions involving charge, spin and lattice degrees of freedom,
 but while many theoretical models for spin and charge stripes have been proposed we
 are missing theoretical models describing the complex lattice striped modulations in doped perovskites.
 In this work we show that two different stripes can be represented as a superposition of a pair of stripes, $U(\theta _n)$  and $D(\theta _n)$,  characterized by perovskite tilts where one of the pair is rotated in relation to the other partner by an angle $ \Delta{\theta _n} = \pi /2 $.
The spatial distribution of the $U$  and $D$  stripes is reduced to all possible maps in
the well-known mathematical four-color theorem. Both the periodic striped puddles
 and random structures can be represented by using $planar$ graphs with a chromatic number $\chi \leqslant 4$.
To observe the colors in mapping experiments, it is necessary to recover variously
oriented tilting effects from the replica. It is established that there is an interplay
between the annihilation/creation of new stripes and ordering/disordering tilts in relation to the $\theta _n$  angle in the CuO$_2$
plane, where the characteristic shape of the stripes coincides with the tilting-ordered regions.
 By their origin, the boundaries between the stripes should be atomically sharp.

\end{abstract}

\maketitle


\section{\label{sec:intr}Introduction\\}

A characteristic feature of perovskite structures ABO$_3$, known as Glazer's systems, is a variety of  tilting effects.~\cite{glazer1972classification,Glazer_2011} The latter can be caused
by external factors, for example, by pressure,~\cite{xiang2017rules} by temperature,~\cite{zhou2004high}
by a discrepancy between the  lattice parameters of a substrate and the material~\cite{Vailionis_2015} or by
the misfit strain between different layers  in complex materials made of stacks of atomic
layers.~\cite{bianconi2000strain,Agrestini_etal2003,Poccia_etal2010,harris2018charge}

The unique functionality of several quantum complex materials with perovskite structure,
like cuprates,~\cite{bussmann2017high,egami2017alex,benedek1993phase,sigmund2012phase,keller2010local,bianconi2001quantum} manganites~\cite{dagotto2001colossal,burgy2004relevance} and
bismuthates~\cite{giraldo2013inhomogeneous}
can be tuned by atomic substitutions, tolerance factor, misfit strain and pressure which control local structural tilts,
bond disproportionation and nanoscale phase separation. A similar complex lattice fluctuations have been identified by EXAFS in BaPb$_{1-x}$Bi$_x$O$_3$ perovskite superconductors showing bond disproportionation and correlated tilts.~\cite{menushenkov2013role,menushenkov2016fermi}

While the competition between disorder and superconductivity has been a topic of intense interest for decades
 there are several exemplary systems can be found in the literature that demonstrate how correlated disorder~\cite{imry1975random,kresin2006inhomogeneous,zaanen2010high,littlewood2011superconductivity,carlson2015condensed} manifested through doping~\cite{fratini2010scale} vacancies and strain~\cite{ricci2013multiscale} in both two and three dimensions,~\cite{giraldo2015stripe,duan2018appearance} plays an important role in the pairing mechanism~\cite{bianconi2000coexistence,menushenkov2016fermi,sakakibara2012multiorbital} and can favor the enhancement of critical temperature~\cite{bianconi2013superconductor,bianconi2012superconductor,bianconi2012enhancement} toward room temperature.~\cite{bianconi2015lifshitz,bianconi2015superconductivity}

 It is known that cuprates show a universal dome of T$_c$ versus doping,
 little is known about the mechanism beyond the variation of the maximum critical temperature of the dome, T$_{c-max}$, which has been found to be controlled by pressure or strain. Using advanced scanning micro x-ray diffraction the space distribution of inhomogeneity in  the nanoscale has been well understood in cuprates. It has been found that a correlated disorder induced by doping and strain leads to charge puddles creating an inhomogeneous lattice, and superconductivity is the result of percolation in a hyperbolic space through filamentary pathways defined by the puddles~\cite{campi2015inhomogeneity,ricci2015direct} which has been confirmed in iron based superconductors,~\cite{duan2018appearance}
in bismuth sulfide~\cite{athauda2019nanoscale} and it is expected also in oxyselenide~\cite{krzton2014superconductivity}
perovskite superconductors.

Since high-$T_c$ superconductors are superlattices of metallic CuO$_2$  layers,
an appropriate physical variable describing elastic effects is not a tolerance factor for three-dimensional ABO$_3$ systems,
but the superlattice misfit strain  between different layers.~\cite{Forgacs_etal1991, Bak_1989}
Using resonant  Cu K-edge x-ray diffraction~\cite{bianconi1996stripe}, Cu K-edge XANES and  Cu K-edge EXAFS~\cite{bianconi1994instability,missori1994evidence} it was shown that  the local structure of the CuO$_2$ plane of doped Bi$_2$Sr$_2$CaCu$_2$O$_{8+y}$ (called BSCCO or Bi2212)~\cite{bianconi1994instability,missori1994evidence, bianconi1996stripe,saini1998local}
 and of doped La$_{2-x}$Sr$_x$CuO$_{4+y}$ (called LSCO or La124) ~\cite{Bianconi_etal1996,lanzara1996temperature,saini2003different,saini2003temperature} perovskite families show a periodic anharmonic incommensurate lattice modulation described as a particular nanoscale phase separation
 due to the formation of alternate stripes of
 undistorted $U$ stripes and distorted $D$ stripes with different tilts (see Fig.\ref{fig:1}) and disproportionated Cu-O bonds in cuprates.~\cite{Bianconi_etal1996}
For the superlattice of CuO$_2$ atomic layers separated by rocksalt LaO spacer layers in the hole-doped layered perovskite
family La$_2$CuO$_4$ (LCO) the relevant physical quantity is the misfit strain between the active [CuO$_2$] and the spacer rocksalt  [La$_2$O$_2+y$] layers.
For the BSCCO superlattice of  perovskite bilayers CaCu$_2$O$_4$  separated by spacer layers  [Sr$_2$O$_2$] and  [Bi$_2$O$_2$]  the relevant physical quantity is the misfit strain between the active CuO$_2$ and the spacer [Bi$_2$O$_2$]  layers. 


Recent  experimental studies using scanning tunneling conductivity (STM) over a wide range of temperatures and doping~\cite{Zhao_etal2019} evidence that there is a clear connection between the stripe structure and electronic structure of Cu oxides supported by the fact that the pseudogap state and the stripe structure detected by EXAFS~\cite{missori1994evidence,lanzara1996temperature,saini2003temperature} and scanning micro x-ray diffraction \cite{campi2015inhomogeneity} have a similar temperature region of existence.

\section{\label{sec:II} Group $G\left( \alpha  \right)$ of stripe pairs \\}

\begin{figure}
\includegraphics{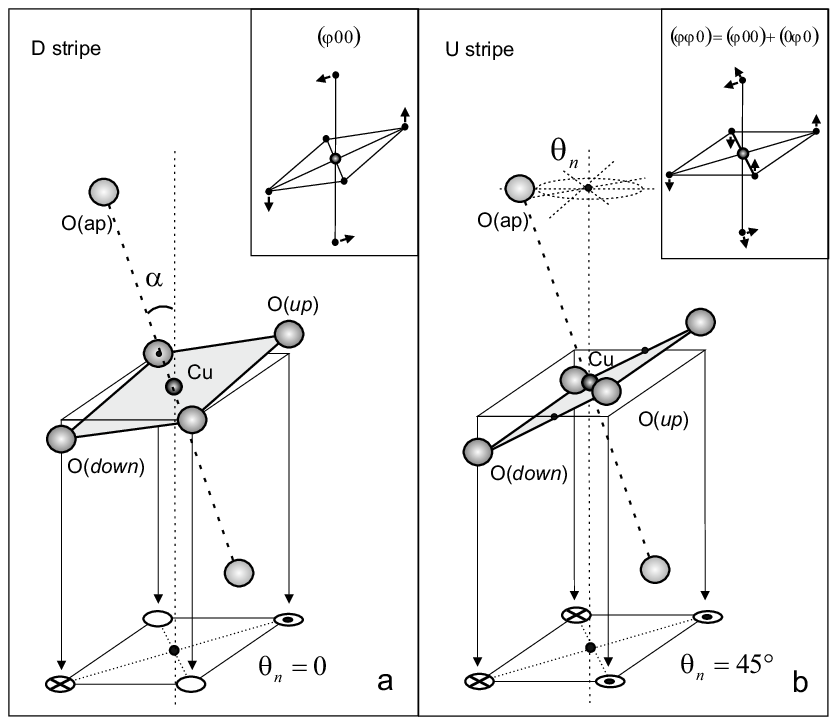}
\caption{View of the distorted   octahedra of the $D$ (a) and $U$ (b)  stripes with different tilting effects in LSCO.~\cite{Bianconi_etal1996} The planar projections in the lower part of the figure corresponds to the positions of oxygen atoms in the CuO$_4$ planar complex. The insets show the tilting effects in accordance with Glazer's  notations.~\cite{Glazer_2011}}
\label{fig:1}
\end{figure}

The research was inspired by the study presented in the work,~\cite{Zhang_etal2018} where hidden symmetries were investigated  in the stripe structure of high-temperature superconductors.
Here, we focus on the tilting effects as a required attribute of observed stripes.~\cite{Bianconi_etal1996}
Based on the large isotope effect in high-T$_c$ superconductors with the stripe nanostructure,~\cite{sasagawa2005oxygen} we tried to study the phonon nature of the stripe structure in high-Tc cuprates, and to identify the stripes according to the type of their tilting effects.  For example, the $U$ and $D$  stripes in Fig.\ref{fig:1} show the $\left( {a^-a^-a^0} \right)$  and   $\left( {a^-a^0a^0} \right)$ tilts in Glazer's  notations.~\cite{Glazer_2011}
 Unfortunately, in these notations it is impossible to classify the stripes rotated about the axis $z$  by different angles $\theta_n$. Therefore we introduce new notations in which an orientation of the tilting angle $\alpha$ can vary within a single Glazer's notation. To do this, we replace $a$ by the angle $\pm \alpha$, and omit the top indices $+$ and $-$ as they do not change under the action of rotation about the axis $z$.
Then, all the possible $U$ and $D$  stripes: $D\left( 0 \right) = \left( {0,\alpha, 0} \right)$, $U\left( 45^\circ \right) = \left( {\alpha,\alpha,0} \right)$, $D\left( {90^\circ } \right) = \left( {\alpha,0,0} \right)$, $U\left( {135^\circ } \right) = \left( {\alpha, - \alpha,0} \right)$, $D\left( {180^\circ } \right) = \left( {0, - \alpha,0} \right)$, $U\left( {225^\circ } \right) = \left( { - \alpha, - \alpha,0} \right)$, $D\left( {270^\circ } \right) = \left( { - \alpha,0,0} \right)$, $U\left( {315^\circ } \right) = \left( { - \alpha,\alpha,0} \right)$ form a group of a complex root of unity $\sqrt[g]{{{e^{\left( {2\pi i} \right)}}}}$, where $g=8$. This is the symmetry group of an isolated CuO$_6$ octahedron with respect to its rotation by an angle $\theta_n=n(\pi/4)$.
 In the LCO and LSCO layered materials, this group reduces to the two  $U\left( {{\theta _n}} \right)$ and $D\left( {{\theta _n}} \right)$  groups because of different orientation of the CuO$_4$ planar complex relative to the rocksalt spacer layers.

\begin{figure}
\includegraphics{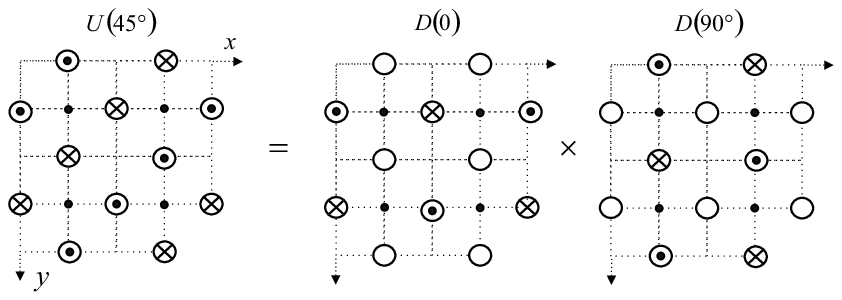}
\caption{ $U$ type structure in the undoped orthorhombic LCO  is presented as a superposition of two $D$  tetragonal structures  rotated one relative to the other by an angle $\pi/2$}
\label{fig:2}
\end{figure}

In accordance with the study,~\cite{Pickett_1989} the undoped LCO has the orthorhombic A$_{bma}$ (D$^{18}_{2h}$) structure at a temperature below 500K, where the tilts of the perovskite octahedra are ordered into a $(a^-a^-a^0)$ sequence. The latter can formally be presented here as a superposition   $\left( {a^-a^-a^0} \right) = \left( {a^-a^0a^0} \right) + \left( {a^0a^-a^0} \right)$, and we further consider the mono $U$ stripe  structure of the undoped LCO as the confinement of  a pair of tetragonal structures  rotated relative to each other by an angle  $\Delta \theta_n  = {\pi  \mathord{\left/
 {\vphantom {\pi  2}} \right.
 \kern-\nulldelimiterspace} 2}$ (Fig.\ref{fig:2}). As a consequence, the number of stripes, in the doped LSCO, is reduced to one species ($U$ stripes) with extensive overlapping areas ($D$ stripes) between them.

 To study the nature of the spatial stripe distribution we take into account that the $U\left( {{\theta _n}} \right)$  and $D\left( {{\theta _n}} \right)$  structures have a common origin and differ only by orientation (i.e. $\theta_n$) of the tilting angle $\alpha$.  Let's collect the $U(\theta_n)$ and $D(\theta_n)$   stripes  into a new group  $G\left( \alpha\right)$ built from the stripes rotated relative to each other by an angle ${\theta _n} = n\left( {\pi /4} \right)$
\begin{eqnarray}
U\left( 45^\circ \right) = D\left( 0 \right) \times D\left( {90^\circ } \right) = \left( {0,\alpha,0} \right) + \left( {\alpha,0,0} \right)\nonumber\\
D\left( {180^\circ } \right) = D{\left( 0 \right)^2} = \left( {0,\alpha,0} \right) + \left( {0,\alpha,0} \right) \nonumber \\
O = D\left( 0 \right) \times D\left( {180^\circ } \right) = \left( {0,\alpha,0} \right) + \left( {0, - \alpha,0} \right)
 \label{eq:1}
\end{eqnarray}
and etc. Following the superposition for  the $U\left( {{\theta _n}} \right)$ and $D\left( {{\theta _n}} \right)$   stripes, we derive the rules of group multiplication for the stripes: $U\left(45^\circ \right)$, $U\left( {135^\circ } \right)$, $U\left( {225^\circ } \right)$, $U\left( {315^\circ } \right)$   and $D\left( 0 \right)$, $D\left( {90^\circ } \right)$, $D\left( {180^\circ } \right)$,   $D\left( {270^\circ } \right)$ (see Tab.1) , where the  group unit $O$ refers to the untilted prototype in which there are no tilts  at all. The group $G\left( \alpha  \right)$ of the stripe pairs presented in Tab.1  appears if we assume the replacement $\left( {2\alpha } \right) \leftrightarrow  - \alpha$  in the squared group element $D{\left( 0 \right)^2} = \left( {0,\alpha,0} \right) + \left( {0,\alpha,0} \right) = \left( {0,2\alpha,0} \right) = D\left( {180^\circ } \right)$ , i.e. there can be only one specific angle $\alpha \sim const$ in the material. In this case, we obtain an abelian group  $G\left( \alpha \right)$ of the ninth order, which is a direct product of any two cyclic subgroups:
\begin{eqnarray}
G &=& {D_1} \times {D_2} = {U_1} \times {U_2} = {D_1} \times {U_1}=  \nonumber \\
&=& {D_2} \times {U_2} ={D_1} \times {U_2} = {D_2} \times {U_1}
 \label{eq:2}
\end{eqnarray}
of the  four possible:
\begin{eqnarray}
{D_1} = D\left( 0 \right),D\left( {180^\circ } \right),O; {U_1} = U\left( 45^\circ \right),U\left( {225^\circ } \right),O; \nonumber \\
{D_2} = D\left( {90^\circ } \right),D\left( {270^\circ } \right),O; {U_2} = U\left( {135^\circ } \right),U\left( {315^\circ } \right),O \nonumber \\
\label{eq:3}
\end{eqnarray}

In contrast to the observed orthorhombic symmetry with the $(a^-,a^-,a^0)$ tilting effects in the undoped LCO cuprates,~\cite{Pickett_1989} described only by the subgroup ${U_1}$  (or ${U_2}$ ) while in the doped cuprates LSCO, the effects $(a^-a^-a^0)$ and $(a^-a^0a^0)$, $(a^0a^-a^0)$ from   the direct product ${U_1} \times {U_2}$ might be observed.~\cite{Bianconi_etal1996}
\begin{table*}
\caption {The multiplication table for the ninth order abelian group $G\left( \varphi  \right)$, where the group unit is an untilted prototype $O$}
\label{tab:1}
\begin{ruledtabular}
\begin{tabular}{cccccccccc}
 \footnotesize Group elements             &\footnotesize $D(0)$            &\footnotesize $D(90^\circ)$ &\footnotesize $D(180^\circ)$    &\footnotesize $D(270^\circ)$   &\footnotesize  $U(45^\circ)$ &\footnotesize  $U(135^\circ)$ &\footnotesize  $U(225^\circ)$ &\footnotesize $U(315^\circ)$ &\footnotesize $O$\\
\hline

\footnotesize $D(0)$ &\footnotesize  $D(180^\circ)$     &\footnotesize   $U(45^\circ)$  &\footnotesize $O$ &\footnotesize $U(315^\circ)$ &\footnotesize $U(135^\circ)$ &\footnotesize $D(90^\circ)$ &\footnotesize $D(270^\circ)$ &\footnotesize $U(225^\circ)$ &\footnotesize $D(0)$\\

\footnotesize $D(90^\circ)$ &\footnotesize $U(45^\circ)$  &\footnotesize $D(270^\circ)$ &\footnotesize $U(135^\circ)$ &\footnotesize $O$ &\footnotesize $U(315^\circ)$ &\footnotesize $U(225^\circ)$ &\footnotesize $D(180^\circ)$ &\footnotesize $D(0)$ &\footnotesize $D(90^\circ)$\\

\footnotesize $D(180^\circ)$ &\footnotesize $O$ &\footnotesize $U(135^\circ)$ &\footnotesize $D(0)$ &\footnotesize $U(225^\circ)$ &\footnotesize $D(90^\circ)$ &\footnotesize $U(0)$ &\footnotesize $U(315^\circ)$ &\footnotesize $D(270^\circ)$ &\footnotesize $D(180^\circ)$\\

\footnotesize $D(270^\circ)$ &\footnotesize $U(315^\circ)$ &\footnotesize $O$ &\footnotesize $U(225^\circ)$ &\footnotesize $D(90^\circ)$ &\footnotesize $D(0)$ &\footnotesize $D(180^\circ)$ &\footnotesize $U(135^\circ)$ &\footnotesize $U(0)$ &\footnotesize $D(270^\circ)$\\

\footnotesize $U(45^\circ)$ &\footnotesize $U(135^\circ)$ &\footnotesize $U(315^\circ)$ &\footnotesize $D(90^\circ)$ &\footnotesize $D(0)$ &\footnotesize $U(225^\circ)$ &\footnotesize $D(270^\circ)$ &\footnotesize $O$ &\footnotesize $D(180^\circ)$ &\footnotesize $U(0)$\\

\footnotesize $U(135^\circ)$ &\footnotesize $D(90^\circ)$ &\footnotesize $U(225^\circ)$ &\footnotesize $U(0)$ &\footnotesize $D(180^\circ)$ &\footnotesize $D(270^\circ)$ &\footnotesize $U(315^\circ)$ &\footnotesize $D(0)$ &\footnotesize $O$ &\footnotesize $U(135^\circ)$\\

\footnotesize $U(225^\circ)$ &\footnotesize $D(270^\circ)$ &\footnotesize $D(180^\circ)$ &\footnotesize $U(315^\circ)$ &\footnotesize $U(135^\circ)$ &\footnotesize $O$ &\footnotesize $D(0)$ &\footnotesize $U(0)$ &\footnotesize $D(90^\circ)$ &\footnotesize $U(225^\circ)$\\

\footnotesize $U(315^\circ)$ &\footnotesize $U(225^\circ)$ &\footnotesize $D(0)$ &\footnotesize $D(270^\circ)$ &\footnotesize $U(0)$ &\footnotesize $D(180^\circ)$ &\footnotesize $O$ &\footnotesize $D(90^\circ)$ &\footnotesize $U(135^\circ)$ &\footnotesize $U(315^\circ)$\\

\footnotesize    \footnotesize $O$         &\footnotesize $D(0)$            &\footnotesize $D(90^\circ)$ &\footnotesize $D(180^\circ)$    &\footnotesize $D(270^\circ)$   &\footnotesize  $U(45^\circ)$ &\footnotesize  $U(135^\circ)$ &\footnotesize  $U(225^\circ)$ &\footnotesize $U(315^\circ)$ &\footnotesize $O$\\
\end{tabular}
\end{ruledtabular}
\end{table*}
The irreducible representations of the group $G(\alpha)$ are homomorphisms from this group into a group of three elements: 1, $e^{2\pi i/3}$, $e^{4\pi i/3}$, that is, a third degree root of unity. The homomorphism is completely defined by how it acts on the generators of a group. In our case, these are two elements which do not belong to the same subgroup, for example $D(0)$ and $U(0)$. If $D(0)$ is mapped to $e^{2\pi i/3}$, and $U(0)$ - to $e^{4\pi i/3}$, then $U(135)=D(0)U(0)$  is mapped to 1, and similarly, for all other elements of the group. The three variants of the homomorphism for each of the pair of generators of the group result in the nine different representations, and the latter are numbered by pairs of indices $\{i,j\}$, where $i(j)=1,2,3$.

Since the active CuO$_2$ plane does not contain any untilted prototypes, there is no neighborhood of  $U$ stripes rotated by $180^\circ$  relative to each other. All of them are separated by the $U\left( {{\theta _n}} \right)$  stripes with ${\Delta \theta _n} = 90^\circ$ (or ${\Delta \theta _n} = 270^\circ$ ), which have an overlapping region ($D\left( {{\theta _n}} \right)$ stripes). For example,  $U\left( 45^\circ \right)$/ $D\left( {270^\circ } \right)$/ $U\left( {135^\circ } \right)$/ $D\left( 0 \right)$/ $U\left( {225^\circ } \right)$/ $D\left( {90^\circ } \right)$/ $U\left( {315^\circ } \right)$/ $D\left( {180^\circ } \right)$/ $U\left( 45^\circ \right)$, or $U\left( 45^\circ \right)$/ $D\left( {270^\circ } \right)$/ $U\left( {135^\circ } \right)$/ $D\left( {270^\circ} \right)$/ $U\left( 45^\circ \right)$/$D\left( {270^\circ } \right)$/ $U\left( {135^\circ } \right)$   with a period of the stripe structure less than the order of the group $G(\alpha)$.

\begin{figure}
\includegraphics{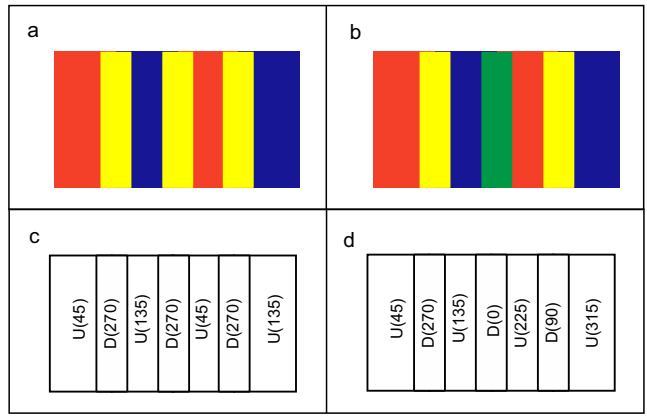}
\caption{(Color online) A color map with the chromatic number $\chi=2$ for the  unidirectional charge-stripe order}
\label{fig:3}
\end{figure}

\begin{figure}
\includegraphics{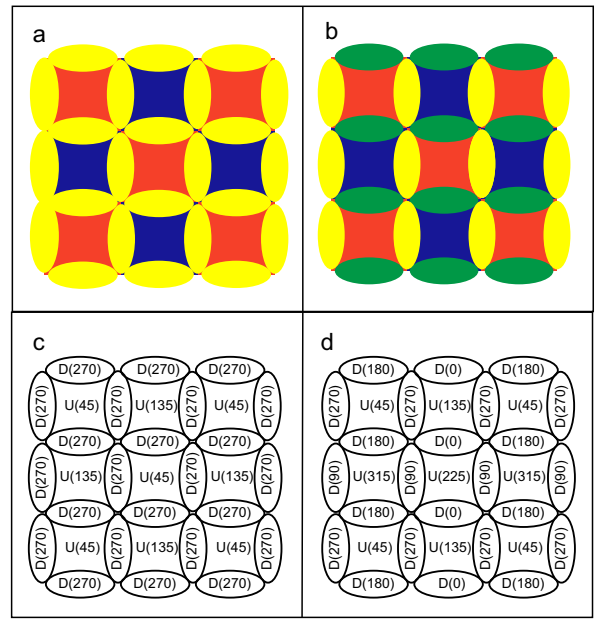}
\caption{(Color online) Color maps with the chromatic number $\chi=3$ for the chessboard electronic crystal.}
\label{fig:4}
\end{figure}

The spatial distribution of the stripes in Fig.\ref{fig:3} and Fig.\ref{fig:4} directly follows from the well-known four color theorem on the map.~\cite{Biggs_etal1986}
In four colors, $U\left( 45^\circ \right)$, $U\left( {225^\circ } \right)$ -red ($R$),
$U\left( {135^\circ } \right)$, $U\left( {315^\circ } \right)$ -blue ($B$), and also
$D\left( 0 \right)$, $D\left( {180^\circ } \right)$ -green ($G$),
$D\left( {90^\circ } \right)$, $D\left( {270^\circ } \right)$ - yellow ($Y$) we can always color an arbitrary map on the plane
.~\cite{Biggs_etal1986} The four colors: $R$, $G$, $B$, $Y$ are just four subgroups:
${D_1}$, ${D_2}$  and ${U_1}$, ${U_2}$, from which the untilted prototypes $O$  have been removed.
For example, the unidirectional charge stripe order and a chessboard electronic crystal~\cite{okamoto2012spontaneous,seibold2007checkerboard} can correspond to one of the color maps: $R$/$Y$/$B$/$Y$/$R$, $R$/ $Y$/$B$/$G$/$R$/$Y$/$B$ in Fig.\ref{fig:3} and Fig.\ref{fig:4} respectively.

Any stripe structures, including a random structure are more convenient for representation using planar graphs with a chromatic number  $\chi \leq 4$.~\cite{Biggs_etal1986}  In accordance with the group $G\left( \alpha \right)$,
 all the maps are symmetric with respect to the replacement of $U$   by $D$  and vice versa.

Recently, the four color theorem has been used to interpret the domain topology of the magnetic material Fe$_x$TaS${_2}$.~\cite{Yoichi_etal21012}

\section{Conclusions}

1. The overlapping between the  stripes may be an alternative for Bloch and Neel type of domain walls in oxides~\cite{ Salia_etal2017} at the subatomic scale. The overlapping looks like the $D$  stripes, whose properties follow from the group multiplication in Tab.1. In this group, the  overlapping areas of the $D$  and $U$ stripes   are equal, and  the untilted prototype  $O$ with $\alpha=0$ is the group unit.  In the undoped LCO  cuprate, the structure of one large   $U$ stripe is observed, and  the material is at the state with  $(a^-a^-a^0)$ tilting effects. Under doping, the disappearance of the long-range stripe order is accompanied with the generation of the $D$  stripes  in the region of the immediate boundary between the $U\left( {{\theta _n}} \right)$ stripes oriented at different angles ($\Delta {\theta _n} =  {\pi /2} $)   relative to each other.
The spatial distribution of the $D\left( {{\theta _n}} \right)$ and $U\left( {{\theta _n}} \right)$  stripes is reduced to possible maps in the four color problem for the CuO$_4$ plane, where the color represents the $D\left( {{\theta _n}} \right)$  or $U\left( {{\theta _n}} \right)$  stripes combined into four subgroups of the group $G(\alpha)$. Any observed spatial distributions, namely, the unidirectional charge-stripe order, chessboard electronic crystal or totally random structures can be represented by planar graphs with different chromatic numbers, where  $2 \leqslant \chi \leqslant 3$  and $\chi = 4$,   respectively.   $\chi = 1$ at the long-range stripe ordering in the undoped material LCO.
By their origin, the boundaries between the $U$   and $D$   stripes  should atomically be sharp.
Moreover, in layered materials the reduced symmetry of the CuO$_4$ square in the  $D$ stripes agrees well with the symmetry of the local ${b_{1g}}$  vibrations that are active in the $\left( {{}^2{a_1} + {}^2{b_1}} \right) \otimes \left( {{b_{1g}} + {a_{1g}}} \right)$  JT effect.~\cite{Bersuker_etal1992,Bersuker_etal1989}. However, unlike the usual local  ${b_{1g}}$ mode,
there no contribution to the multimode JT effect from the tilting mode in the CuO$_6$  octahedron in the isotropic environment.
In the $U$ stripes, the local symmetry of all four plane oxygens is identical even in layered materials. We expect the inhomogeneous distribution of electron density over the $U$  and $D$  stripes since the JT energy depends on the electron concentration only in the $D$  stripes. To observe the colors using STM spectroscopy, it is necessary to recover variously oriented tilting effects from the STM charge replica of the CuO$_2$ plane.
Note, compatible types of cooperative JT and tilting effects within a single structure were investigated in the study,~\cite{Carpenter_etal2009} and an interplay between the tilting and JT distortions was studied previously in LaMnO$_3$, YVO$_3$, and YTiO$_3$.~\cite{Mizokawa_etal1999}

2. What is the physical meaning of the group $G(\alpha)$?  It is not a symmetry group of the lattice
Hamiltonian since the group elements $U\left( {{\theta _n}} \right)$  and $D\left( {{\theta _n}} \right)$
do not reproduce any initial stripe distribution.
The group $G(\alpha)$   is just the symmetry of a pair of interacting stripes.
By analogy with the Woodward-Hoffmann's rules for orbital symmetry conservation in chemical reactions,~\cite{Woodward_etal1971}
 we can associate the initial and final reagents in a reaction with the stripes
 $U\left( {{\theta _n}} \right)$  and $D\left( {{\theta _n}} \right)$, respectively.
In this reversible "reaction", the orbital symmetry changes, but the tilting angle $\alpha$   remains a constant.
Due to the group, we have established the relationship between the annihilation/creation of new stripes and ordering/disordering process of the tilts in the CuO$_2$   plane, where the characteristic shape of the stripes coincides with the regions ordered by the $\theta _n$ angle.
\begin{acknowledgments}
One of the authors (VAG) is grateful to RFBR, the Government of the Krasnoyarsk Region and Krasnoyarsk Regional Fund of Science for the research grant No.18-42-240017
\end{acknowledgments}
\bibliography{ShortPRB3_2019}
\end{document}